\documentstyle[aaspp4]{article}
\begin{document}

\title{Small-scale Interaction of Turbulence with \\ Thermonuclear Flames
in Type Ia Supernovae}

\author{J. C. Niemeyer}
\affil{University of Chicago, Deptartment of Astronomy and Astrophysics,\\
5640 S. Ellis Avenue, Chicago, IL 60637,}
\author{W. K. Bushe and G. R. Ruetsch}
\affil{Center for Turbulence Research, Stanford University,
Stanford, CA 94305}

\begin{abstract}
Microscopic turbulence-flame interactions of thermonuclear
fusion flames occuring in Type Ia Supernovae were studied by means of
incompressible direct numerical simulations with a highly simplified
flame description. The 
flame is treated as a single diffusive scalar field with a nonlinear
source term.  It is characterized by its Prandtl number, $Pr \ll 1$,
and laminar flame speed, $S_{\rm L}$. We find that if $S_{\rm L} \ge
u'$, where $u'$ is the $rms$ amplitude of turbulent velocity
fluctuations, the local flame propagation speed does not significantly
deviate from $S_{\rm L}$ even in the presence of velocity fluctuations
on scales below the laminar flame thickness. This result is
interpreted in the context of subgrid-scale modeling of supernova
explosions and the mechanism for deflagration-detonation-transitions.
\end{abstract}


\section{Introduction}

The thermonuclear explosion of a Chandrasekhar mass C+O white dwarf is
presently the most promising candidate to explain the majority of
Type Ia Supernova (SN Ia) events (H\"oflich et al.~1996). However,
the complex phenomenology of turbulent thermonuclear flames and
deflagration-detonation-transitions (DDTs) renders a self-consistent
description of the explosion mechanism extremely difficult
(Khokhlov 1995, Niemeyer et al.~1996, Niemeyer \& Woosley 1997). The
open questions can be broadly classified as macroscopic ones,
pertaining to the global structure of the flame front and the
buoyancy-driven production of turbulence, and microscopic ones
including turbulence-flame interactions on scales of the flame
thickness and pre-conditioning for DDT. In this work, first results of an
investigation of the latter will be presented, obtained from
direct numerical simulations of a simplified flame model coupled to a
three-dimensional incompressible turbulent flow.

Based on the observational evidence of intermediate elements in SN Ia
spectra, detonations can be ruled out as the initial combustion mode
after onset of the thermonuclear runaway, as they would predict the
complete incineration of the
white dwarf to iron group nuclei. Deflagrations, on the other hand,
are hydrodynamically unstable to both flame intrinsic
(Landau-Darrieus) and buoyancy-driven (Rayleigh-Taylor, RT)
instabilities. While the former is stabilized in the nonlinear regime,
the latter produces a growing, fully turbulent RT-mixing region of hot
burning products and cold ``fuel'', separated by the thin
thermonuclear flame. Driven predominantly by the shear flow
surrounding buoyant
large-scale bubbles, turbulent velocity fluctuations cascade down to
the Kolmogorov scale $l_{\rm k}$, which may, under certain
conditions, be smaller than the laminar flame thickness (Section
2). This regime  is unknown territory for flame modeling; although it has
been speculated in the supernova literature that the effect of
turbulence on the laminar flame structure is negligible as long as the
velocity fluctuations are sufficiently weak, the existence of
turbulent eddies on scales smaller than the flame thickness --
regardless of their velocity --  is in
conflict with the definition of the ``flamelet regime'' in the
flamelet theory of turbulent combustion (Peters 1984). No numerical or
experimental evidence to confirm and quantify this speculation has
been available so far.

As the explosion proceeds, the turbulence intensity grows while the
flame slows down and thickens as a consequence of the decreasing
material density of the expanding star. After some time, small scale
turbulence must be expected to significantly alter the flame structure
and its local propagation velocity with respect to the laminar
solution. On the other hand, most subgrid-scale models for the
turbulent thermonuclear flame brush in numerical simulations of
supernovae depend crucially on the assumption of a (nearly) laminar
flame structure on small scales (Niemeyer \& Hillebrandt 1995,
Khokhlov 1995, Niemeyer et al.~1996). The intent of this work is to
present a first approach 
to study the regions of validity and the possible breakdown of this
``thermonuclear flamelet'' assumption. Specifically, a modification of
Peters' (1984) flamelet definition suggested by Niemeyer \& Kerstein
(1997) will be tested. 

In addition to the verification of subgrid-scale models, this inquiry
is relevant in the context of DDTs which were suggested to occur
in SN Ia explosions after an initial turbulent deflagration phase (Khokhlov
1991, Woosley \& Weaver 1994). A specific mechanism for DDT in SN Ia explosions
based on strong turbulent straining of the flame front and transition to
the distributed burning regime has been proposed (Niemeyer \& Woosley
1997). The ratio of laminar flame speed to turbulence velocity on the
scale of the flame thickness, $S_{\rm L}/u(\delta)$, where $\delta$ is
the laminar thermal flame thickness, has been suggested as a control
parameter indicating the transition to distributed burning when
$S_{\rm L}/u(\delta) \sim O(1)$ (Niemeyer \& Kerstein 1997). One of
the main results presented below
is that the transition to distributed burning was not observed in the
parameter range ($S_{\rm L}/u(\delta) \ge 0.95$) that we were able to
probe. 

Thermonuclear burning fronts are similar in many ways to
premixed chemical flames. The issues addressed in this work are motivated in the
framework of supernova research, but our results apply
equally well to premixed chemical flames with low Prandtl numbers and
small thermal expansion rates. In order to facilitate numerical
computations, we modeled the flame with a single scalar
reaction-diffusion equation that is advected in a three-dimensional,
driven incompressible turbulent flow. The arguments justifying these
simplifications are outlined in Section 2.

This paper is organized as follows: We shall summarize the most
important parameters and dimensional relations of thermonuclear flames
and buoyancy-driven turbulence in Section 2, followed by a brief
description of the numerical methods employed for this work (Section
3). In Section 4, the results of a series of direct simulations of a
highly simplified flame propagating through a turbulent medium are
discussed and interpreted in the framework of SN Ia modeling. 

\section{Flame properties and model formulation}

The laminar properties of thermonuclear flames in white dwarfs were
investigated in detail by Timmes \& Woosley (1992), including all
relevant nuclear reactions and microscopic transport mechanisms. The authors
found that the laminar flame speed, $S_{\rm L}$, varies between $10^7$
and $10^4$ cm s$^{-1}$ as the density declines from $3 \times 10^9$ to
$\sim 10^7$ g cm$^{-3}$. The thermal flame thickness, $\delta$,
grows from $10^{-5}$ to $1$ cm for the same density
variation. Microscopic transport is dominated entirely by electrons
close to the Fermi energy by virtue of their near-luminal
velocity distribution and large mean-free-paths. As a consequence,
ionic diffusion of nuclei is negligibly small compared with heat
transport and viscosity. Comparing the latter two, one finds typical
values for the Prandtl number of $Pr = \nu/\kappa \approx 10^{-5}
\dots 10^{-4}$, where $\kappa$ and $\nu$ are the thermal diffusivity
and viscosity, respectively
(Nandkumar \& Pethick 1984). Further, partial electron degeneracy in
the burning products limits the density contrast, $\mu = \Delta
\rho/\rho$, between burned and unburned material to very small values,
$\mu \approx 0.1 \dots 0.5$.  

To within reasonable accuracy, one may estimate the magnitude of
large-scale turbulent velocity fluctuations, $u(L)$, from the rise
velocity of buoyant bubbles with diameter $L$, 
$u_{\rm rise} \sim (0.5 \mu g L)^{1/2}$, where $g$ is the
gravitational acceleration. Inserting typical values, $L \approx
10^{7}$ cm, $g \approx 10^8$ cm s$^{-2}$, and $\mu \approx 0.3$, one
finds $u(L) \approx 10^7$ cm s$^{-1}$. For a viscosity of $\nu \approx
1$ cm$^2$ s$^{-1}$ (Nandkumar \& Pethick 1984), this yields the
integral-scale Reynolds number $Re \approx 10^{14}$ and a
characteristic Kolmogorov scale $l_{\rm k} \approx L Re^{-3/4} \approx
10^{-4}$ cm. Hence, it is clear that soon after the onset of the
explosion, turbulent eddies are present on scales smaller than the
laminar flame thickness. In conventional flamelet theory (Peters
1984), the ``flamelet regime''   
is defined based on length-scale arguments alone; that is, if the
characteristic length-scale of the flame is smaller than the Kolmogorov
length, the turbulent flame is said to be in the flamelet regime.
Thus, according to
conventional flamelet theory, the scaling arguments offered here would
clearly indicate that these thermonuclear flames are not in the flamelet
regime.  Therefore, flamelet-based models such as those used in almost
all multidimensional SN Ia simulations would not appear to be
applicable for these flames.  

However, the low Prandtl number of degenerate matter allows a
situation in which the Kolmogorov time
scale, $\tau_{\rm k} \sim l_{\rm k}/u(l_{\rm k}) \sim l_{\rm k}^2/\nu$,
is larger than the reaction time scale $\tau_{\rm r} \sim \dot
w^{-1}$, where $\dot w$ is the fuel consumption rate (Niemeyer \&
Kerstein 1997). This is readily
seen by setting $\tau_{\rm r}$ equal to the diffusion time scale
$\tau_{\rm d} \sim \delta^2/\kappa$ for stationary flames (where
$\kappa$ is the microscopic thermal diffusivity), yielding 
\begin{equation}
{\tau_{\rm k} \over \tau_{\rm r}} = Pr^{-1} \left(l_{\rm
k} \over \delta \right)^2 \,\,.
\end{equation}
Even if the length scale ratio on the $rhs$ is less than unity, the
$lhs$ can be large for a sufficiently small $Pr$. In this case, small
eddies are burned before their motion can appreciably affect the flame
structure. 

An alternative, $Pr$-independent, criterion for flamelet breakdown has
been proposed (Niemeyer \& Kerstein 1997), based on the relative
importance of eddy diffusivity, $\kappa_{\rm e} \sim u(l) l$, and
microscopic heat conductivity on scales $l \le \delta$. As
$\kappa_{\rm e}$ is, in general, a growing function of scale, the
condition $\kappa_{\rm e}(\delta) \le \kappa$ is sufficient and can be
invoked to define the flamelet burning regime. Using the relation
$S_{\rm L}\sim\delta/\tau_{\rm d}$, one finds the more intuitive
formulation $u(\delta) \le S_{\rm L}$. In other words, the flame
structure on scales $\delta$ and below is dominated by heat diffusion
as long as the characteristic velocity associated with eddies of
a length scale the same order as the laminar flame thickness is smaller
than the laminar flame speed. If heat diffusion 
is the only relevant microscopic transport process, the local flame
speed is expected to remain comparable to $S_{\rm L}$ despite the
presence of eddies within the flame.

This paper attempts to establish whether or not
the newly proposed scaling relationship of Niemeyer \& Kerstein is an
appropriate definition of the flamelet regime for thermonuclear flames,
and, more generally, whether or not these thermonuclear flames can be
treated as flamelets in numerical simulations.
In order to be able to efficiently address this question, we make
three assumptions that greatly simplify the problem without violating
the underlying physics.  Firstly, we note that nuclear energy
generation is dominated by 
carbon burning which has a very strong dependence on
temperature ($\dot w \sim T^{21}$). Therefore, the flame dynamics
can be well approximated by a single, diffusive progress variable $c$
that is advected by the fluid and coupled to a strongly nonlinear
source term that mimics nuclear burning. Second, the small value of
$\mu$ suggests that dilatation effects do not play a significant role
and may be neglected for the purpose of this study. This, together
with the small Mach number of turbulent fluctuations on very small
scales, justifies the use of the incompressible Navier-Stokes
equations. Finally, we assume that the effect of the turbulent cascade
from large scales can be adequately modeled by forcing the flow field
on the lowest wavenumbers of the simulation.

\section{Numerical technique}

The code used to simulate the thermonuclear flame used the
pseudo-spectral approach, where derivatives are taken in Fourier space
but non-linear terms are evaluated in real space (Ruetsch \&
Maxey 1991).  The diffusive term is evaluated implicitly, such that
the code provided stable, accurate solutions, even for very small
Prandtl numbers.  All boundary conditions were periodic, and energy
was added at every time step to the lowest wavenumbers by solving a
Langevin equation as described in Eswaran and Pope (1988a, 1988b).
All of the simulations were carried out in a $64^3$ domain, and were
run for several eddy-turnover times so as to obtain statistical
stationarity.

As was mentioned in the previous section, the temperature dependence
of the main reaction participating in thermonuclear flame is roughly
$T^{21}$.  It was found that a source term $\dot w = kc^{21}(1-c)$
(where the $(1-c)$ arises from the dependence of the reaction on
reactant concentration) produced too narrow a reaction zone to be
easily resolved in space in a three-dimensional simulation.  Instead,
it was decided to use a source term of $\dot w = kc^4(1-c)$, which is
still strongly non-linear, but produces a reaction zone that can be
resolved in a practical three-dimensional simulation.

\begin{figure}[t]
\begin{minipage}{70mm}
\epsfxsize=70mm
\epsfbox{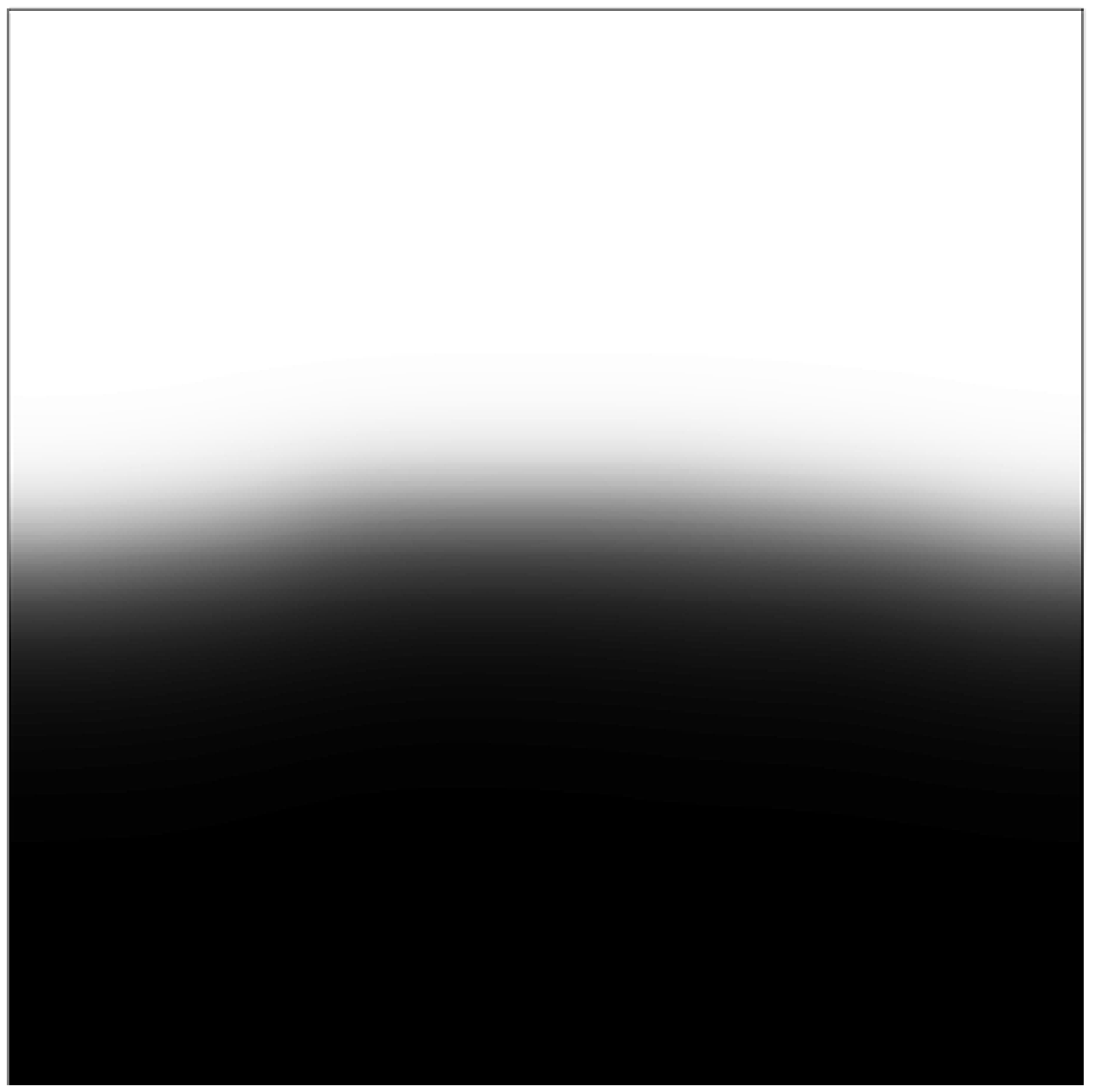}
\caption{\label{f1}Snapshot of the scalar field $c$ for $S_{\rm L}/u' =
11.5$ and $Pr = 0.005$.}
\end{minipage}
\hspace{\fill}
\begin{minipage}{70mm}
\epsfxsize=70mm
\epsfbox{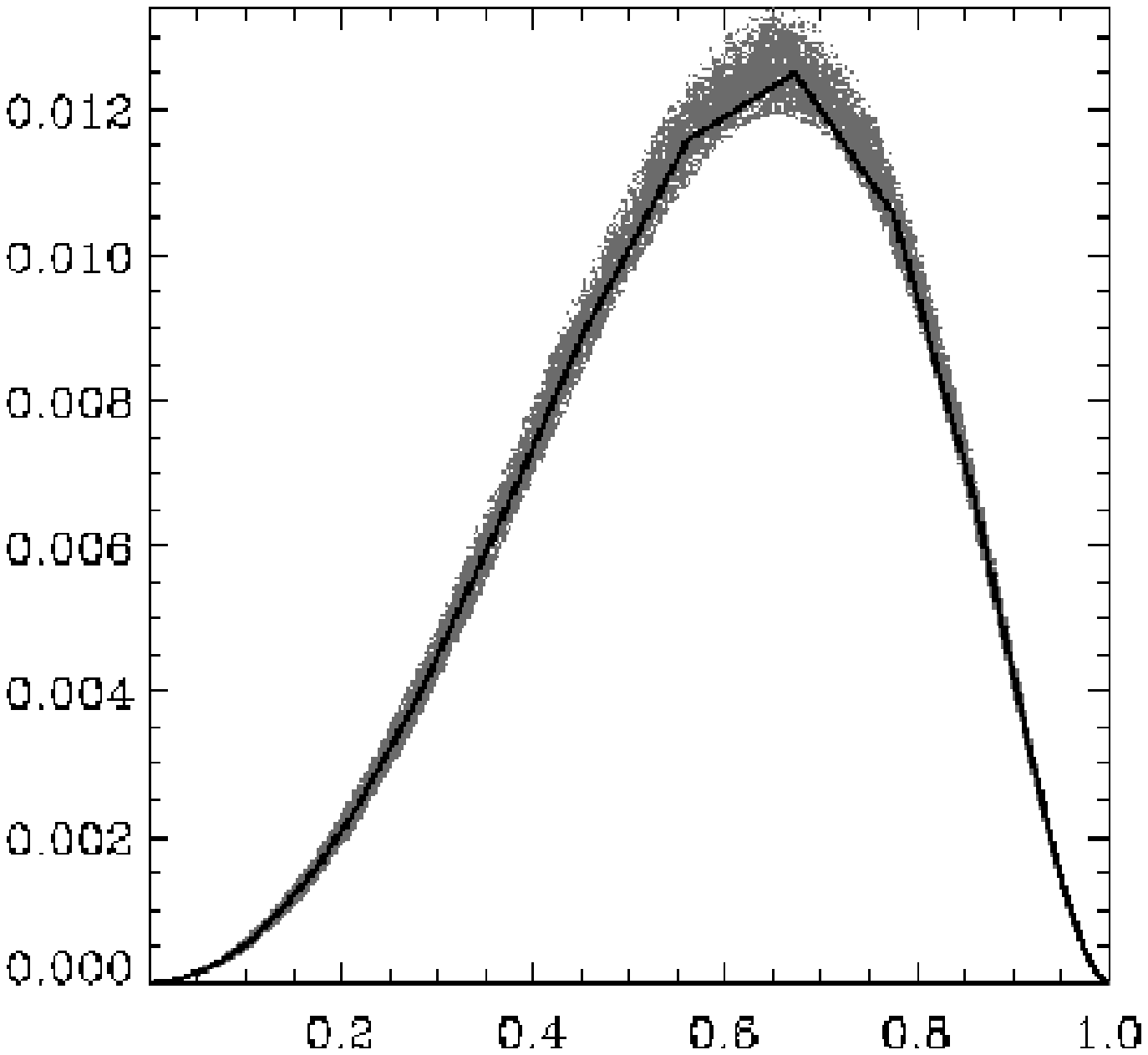}
\caption{\label{f2}Scalar dissipation rate $(\nabla c)^2$ as a
function of $c$ at a fixed time. Superimposed is the line
corresponding to the laminar solution.}
\end{minipage}
\end{figure}
One difficulty that arises in using a pseudo-spectral code to simulate
premixed combustion is that the scalar field -- in this case, the
progress variable -- must be periodic.  This was achieved by separating
the scalar field into two components -- a uniform gradient in the
direction of propagation of the flame was subtracted such that the
remaining field was zero at each end of the periodic box in that
direction.  Thus, where
\begin{equation}
{{\partial c}\over {\partial t}}+u_i{{\partial c}\over {\partial
x_i}}={\cal D}{{\partial^2 c}\over {\partial x_i\partial x_i}}+\dot w
\end{equation}
is the transport equation for the progress variable with constant
properties, if a uniform gradient $\beta$ in the $x_3$ direction (the
direction of propagation of the flame) is subtracted,
\begin{equation}
c=\beta x_3+\theta
\end{equation}
then the transport equation for the periodic fluctuating component
$\theta$ is:
\begin{equation}
{{\partial \theta}\over {\partial t}}+u_i{{\partial \theta}\over
{\partial x_i}}+\beta u_3={\cal D}{{\partial^2 \theta}\over {\partial
x_i\partial x_i}}+\dot w\,\,.
\end{equation}

So long as the reaction zone remained relatively thin and did not
approach the boundaries, $c$ remained bounded between 0 and 1.  In
order to keep the reaction away from the boundaries, the mean velocity
in the direction of propagation was set to the propagation speed of
the flame.  This propagation speed was determined at each time step
from a volume integral of the source term. The need to keep the reaction
away from the boundaries was found to restrict the simulation to a
limited ratio of Prandtl number to $k$ -- the flame speed could not be
significantly lower than $u'$ or wrinkles in the flame would become
too large to be contained in the domain.

\section{Discussion of the results}

\begin{figure}[t]
\begin{minipage}{70mm}
\epsfxsize=70mm
\epsfbox{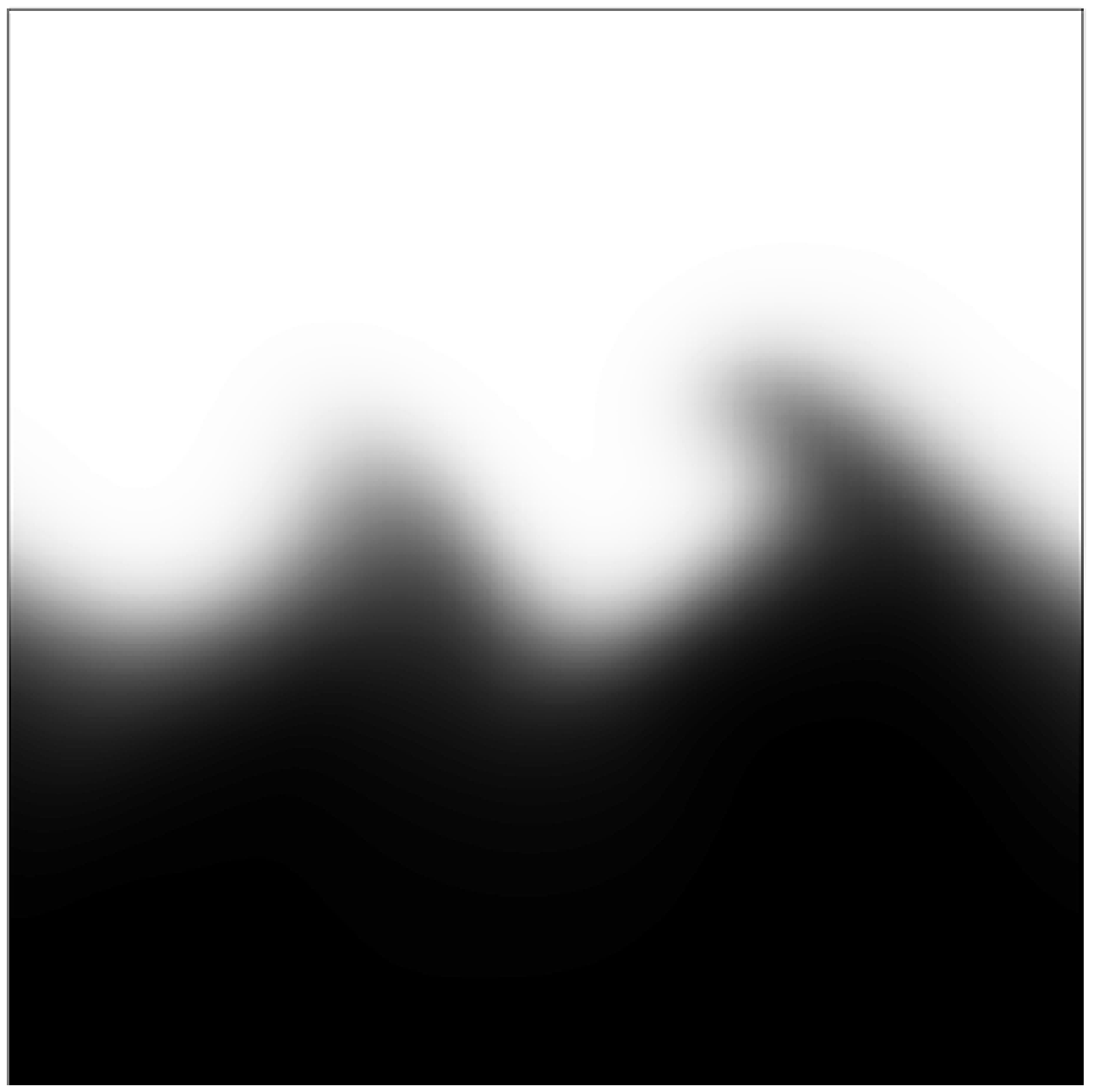}
\caption{\label{f3}Snapshot of the scalar field $c$ for $S_{\rm L}/u' =
1.15$ and $Pr = 0.05$.}
\end{minipage}
\hspace{\fill}
\begin{minipage}{70mm}
\epsfxsize=70mm
\epsfbox{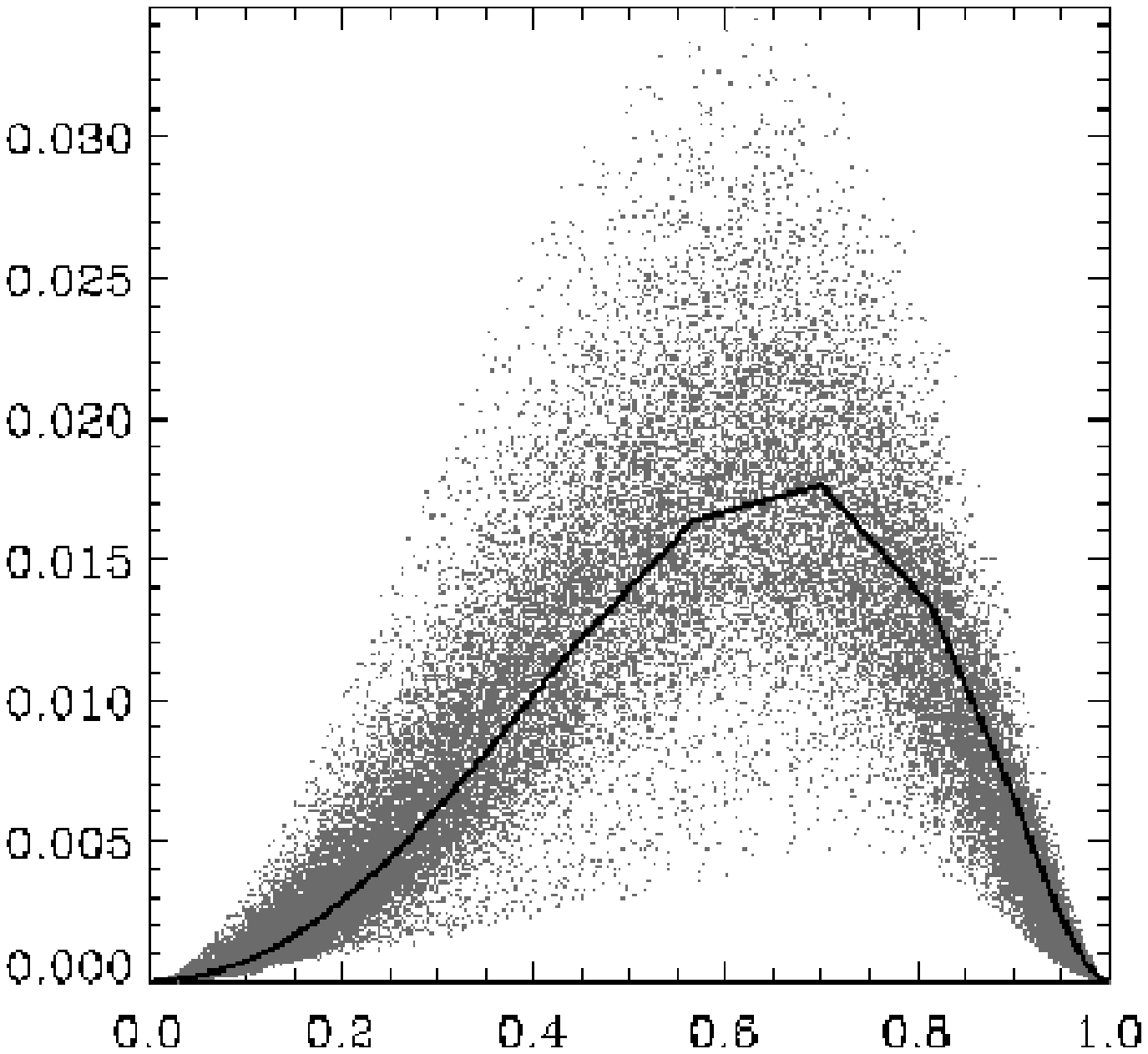}
\caption{\label{f4}Scalar dissipation rate $(\nabla c)^2$ as a
function of $c$ at a fixed time. Superimposed is the line
corresponding to the laminar solution.}
\end{minipage}
\end{figure}

\begin{figure}[t]
\begin{minipage}{70mm}
\epsfxsize=70mm
\epsfbox{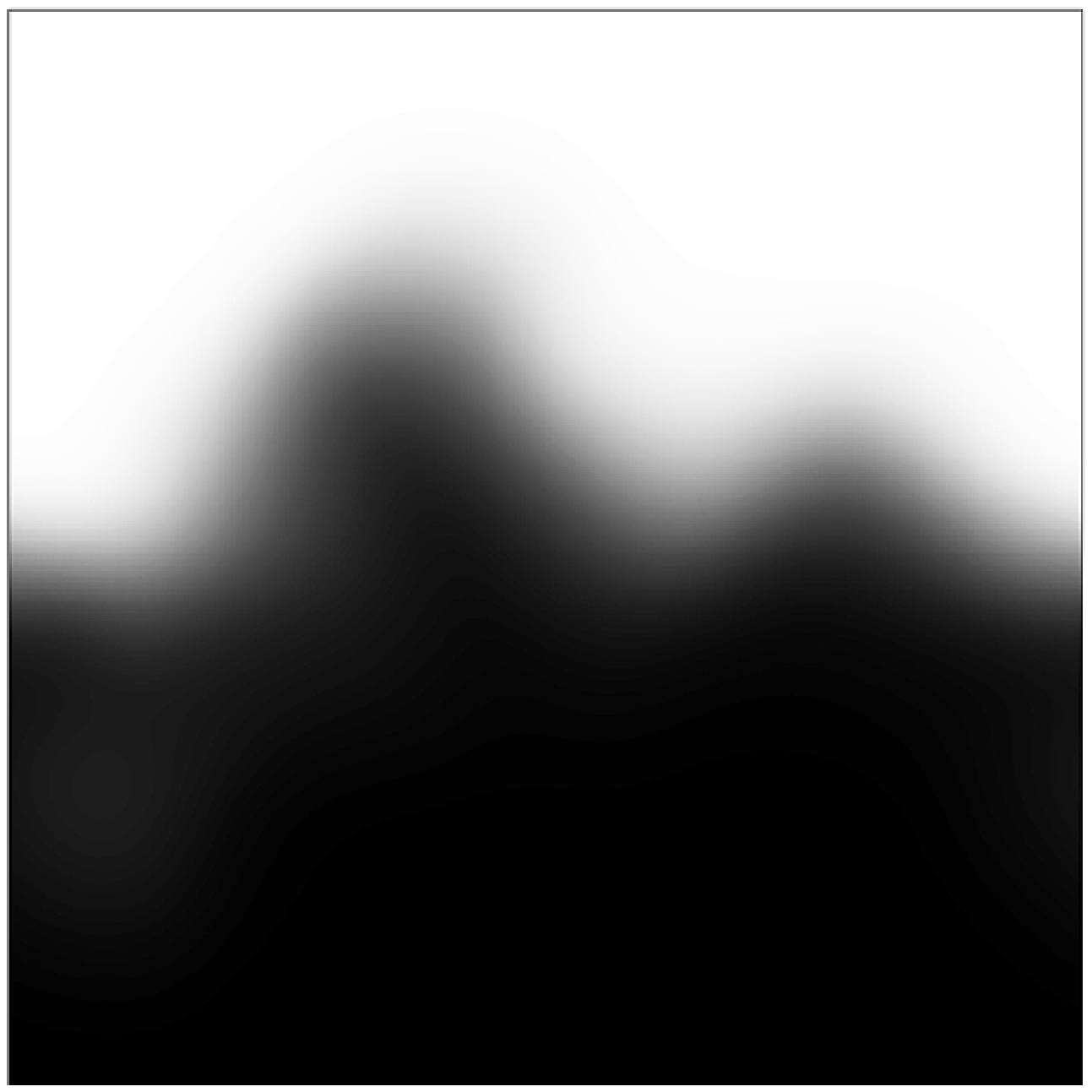}
\caption{\label{f5}Snapshot of the scalar field $c$ for $S_{\rm L}/u' =
0.95$ and $Pr = 0.05$.}
\end{minipage}
\hspace{\fill}
\begin{minipage}{70mm}
\epsfxsize=70mm
\epsfbox{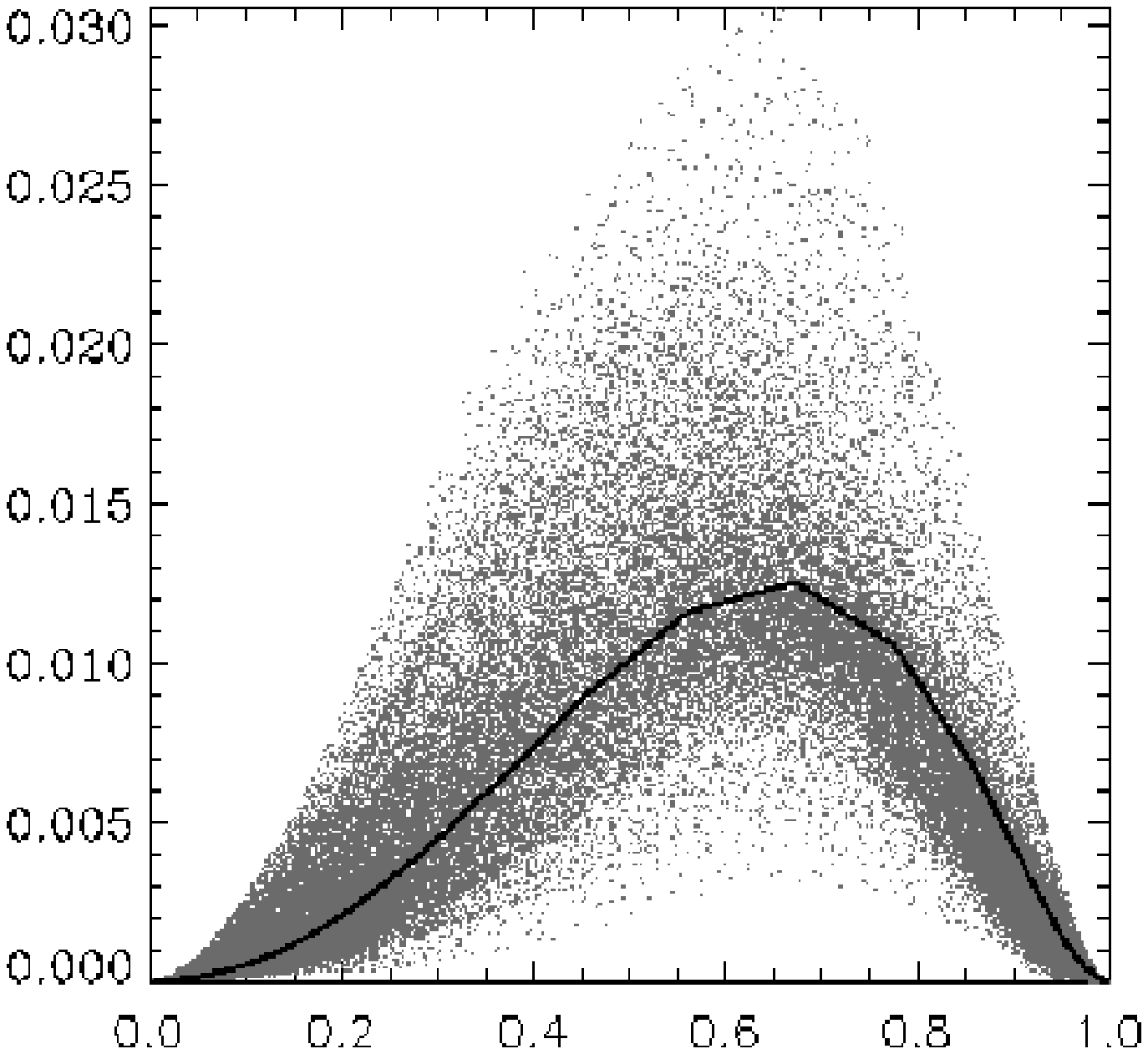}
\caption{\label{f6}Scalar dissipation rate $(\nabla c)^2$ as a
function of $c$ at a fixed time. Superimposed is the line
corresponding to the laminar solution.}
\end{minipage}
\end{figure}
The results of three simulations with varying laminar flame speeds and
Prandtl numbers are illustrated in figures (1) -- (6) (see
figure captions for the model parameters). Note that
$S_{\rm L}/u'$, with the root-mean-square velocity fluctuation $u'$
dominated in the simulation by eddies on the scale of the laminar
flame thickness, corresponds roughly to the parameter $S_{\rm
L}/u(\delta)$  employed in Section (2) to describe the validity of the
flamelet assumption based on dimensional analysis. Therefore, one may
expect noticeable deviations from locally laminar flame propagation
for $S_{\rm L}/u' < 1$. Conversely, the dimensional argument
predicts that changes of the total burning rate are exclusively due to
the growth of the flame surface area by turbulent wrinkling as long as
$S_{\rm L}/u' \ge 1$. 

We define the turbulent flame speed in terms of the volume integral of the
source term, $S_{\rm T} \equiv \Lambda^{-2} \int_V \dot w {\rm d}^3
\lambda$, where $\Lambda$ is the grid length. The wrinkled flame
surface area, $A_{\rm T}$, is measured by triangular discretization of
the $c = 0.5$ isosurface. For the three cases with $S_{\rm L}/u' =
11.5$, $1.15$, and $0.95$ we find $S_{\rm T}/S_{\rm L}$ ($A_{\rm
T}/\Lambda^2$) of 1.008 (1.008), 1.31 (1.27), and 1.51 (1.56),
respectively. Hence, to within 5 \% accuracy the ratio of turbulent
and laminar flame speeds is identical to the increase of the flame
surface area with respect to the laminar surface, implying that the
local flame speed is, on average, equal to $S_{\rm L}$ in all cases.

In conclusion, we confirmed -- within the limitations of our simplified
flame description -- that the local propagation speed of turbulent
low-$Pr$ premixed flames remains equal to $S_{\rm L}$ if $S_{\rm L}
\ge v(\delta)$, even if eddies exist on scales smaller than the flame
thickness. Our results show no indication of a breakdown 
of the flamelet burning regime in the parameter range $S_{\rm
L}/v(\delta) \ge 0.95$ that was studied. Lower values of $S_{\rm
L}/v(\delta)$ were unattainable because large scale flame
wrinkling forced regions with nonvanishing $\dot w$ over the
streamwise grid boundaries, violating the requirement of periodicity
of the non-linear component of the progress variable. This outcome
suggests that the conventional 
definition of the flamelet regime (Peters 1984) which is based on a
length-scale argument alone should be generalized to a time-scale
dependent definition in the sense of Niemeyer \& Kerstein (1997). 

In the framework of supernova modeling, this result helps to formulate
a subgrid-scale model for the turbulent thermonuclear flame brush in
large-scale hydrodynamical simulations. Specifically, it is possible
to estimate $S_{\rm L}/v(\delta)$ from the filtered density and
velocity strain, using an assumed spectrum for the turbulent velocity
cascade. If $S_{\rm L}/v(\delta) \ge 1$, a subgrid-scale model based
purely on the surface increase by turbulent wrinkling can be employed
(Niemeyer \& Hillebrandt 1995). In practice, this is possible for
densities above $\sim 10^7$ g cm$^{-3}$, where most of the explosion
energy is released. For lower densities (in the late stages of the
explosion), relevant for the nucleosynthesis of intermediate mass
elements and a possible deflagration-detonation-transition (Niemeyer
\& Woosley 1997), a more detailed model accounting for small-scale
turbulence flame interactions needs to be developed.

All the currently discussed scenarios for
deflagration-detonation-transitions (DDT) in the late stage of SN Ia
explosions require an earlier transition to distributed or
well-stirred burning in order to allow pre-conditioning of unburned
material. Our results indicate that the flamelet structure of
thermonuclear flames is more robust than previously anticipated, hence
delaying or even preventing the formation of favorable conditions for
DDT during the first expansion phase. A more detailed investigation of this
question, extending the parameter range to lower $S_{\rm L}/v(\delta)$,
is underway (Young, Niemeyer \& Rosner 1999).

\acknowledgements
We would like to thank Joel Ferziger, Nigel Smith, and Dan Haworth for
interesting discussions. JCN wishes to acknowledge the hospitality of
the Center for Turbulence Research where most of this research was
carried out, supported in part by the  
ASCI Center on Astrophysical Thermonuclear Flashes at the University
of Chicago (DOE contract no. B34149).

\end{document}